\documentclass[reprint, amsmath,amssymb, aps, prl,superscriptaddress,email]{revtex4-2}

\usepackage{physics}
\usepackage{graphicx}
\usepackage{svg}
\usepackage{bm}
\usepackage{natbib}
\usepackage{dsfont}
\usepackage{xcolor}
\usepackage{hyperref}
\usepackage{subfiles} 

\begin{document}
\title{Coherent spin-phonon scattering in facilitated Rydberg lattices}
\author{Matteo Magoni}
\affiliation{Institute for Theoretical Physics, University of Innsbruck, Innsbruck 6020, Austria}
\affiliation{Institute for Quantum Optics and Quantum Information, Austrian Academy of Sciences, Innsbruck 6020, Austria}
\affiliation{Institut f\"ur Theoretische Physik, Universit\"at T\"ubingen, Auf der Morgenstelle 14, 72076 T\"ubingen, Germany}
\author{Chris Nill}
\affiliation{Institut f\"ur Theoretische Physik, Universit\"at T\"ubingen, Auf der Morgenstelle 14, 72076 T\"ubingen, Germany}
\affiliation{Institute for Applied Physics, University of Bonn, Wegelerstraße 8, 53115 Bonn, Germany}
\author{Igor Lesanovsky}
\affiliation{Institut f\"ur Theoretische Physik, Universit\"at T\"ubingen, Auf der Morgenstelle 14, 72076 T\"ubingen, Germany}
\affiliation{School of Physics and Astronomy and Centre for the Mathematics and Theoretical Physics of Quantum Non-Equilibrium Systems, The University of Nottingham, Nottingham, NG7 2RD, United Kingdom}
\date{\today}

\begin{abstract}
We investigate the dynamics of a spin system with facilitation constraint that can be studied using Rydberg atoms in arrays of optical tweezer traps. The elementary degrees of freedom of the system are domains of Rydberg excitations that expand ballistically through the lattice. Due to mechanical forces, Rydberg excited atoms are coupled to vibrations within their traps. At zero temperature and large trap depth, it is known that virtually excited lattice vibrations only renormalize the timescale of the ballistic propagation. However, when vibrational excitations are initially present --- i.e., when the external motion of the atoms is prepared in an excited Fock state, coherent state or thermal state --- resonant scattering between spin domain walls and phonons takes place. This coherent and deterministic process, which is free from disorder, leads to a reduction of the power-law exponent characterizing the expansion of spin domains. Furthermore, the spin domain dynamics is sensitive to the coherence properties of the atoms' vibrational state, such as the relative phase of coherently superimposed Fock states. Even for a translationally invariant initial state the latter manifests macroscopically in a phase-sensitive asymmetric expansion.
\end{abstract}

\maketitle
\textbf{Introduction ---}
One of the central topics in quantum many-body physics relates to the study of transport properties of excitations, correlations or energy, which allows to categorize models into different classes~\cite{Bertini_review,Chien2015}. For example, chaotic systems generically feature diffusive transport~\cite{Karrasch_2014,Lux_2014,Gu2017,Blake_2017,Friedman_2020}, while disorder may induce subdiffusive dynamics or even localization~\cite{Basko_2006,Bar_Lev_2015,Imbrie2016,Znidaric_2016}. On the contrary, ballistic transport and diffusive dynamics are typically featured in free and interacting integrable systems~\cite{Sirker_2009,Znidaric_2011,Ilievski_2017,Medenjak_2017,De_Nardis_2018,Gopalakrishnan_2018,Jepsen2020,Dupont2020,Bastianello_2021}, since they are characterized by an extensive number of local conserved charges~\cite{MBL:Review}. An intermediate behavior between diffusive and ballistic dynamics has been observed, also experimentally~\cite{Hild_2014,Scheie2021,Wei_2022,Joshi_2022,Wienand_2023}, in a class of integrable models with certain additional symmetries~\cite{Prosen_nat,Ilievski_2018,De_Nardis_2019,Gopalakrishnan_2019,Ljubotina_2019,Bulchandani_2020}, which appear to lie in the Kardar-Parisi-Zhang universality class~\cite{Kardar_Parisi_Zhang}.

Recently, transport properties have also been studied in quantum systems subject to kinetic constraints~\cite{Lan2018,Pancotti_2020,Yang_2022,Feldmeier_2022,Ljubotina_2023,Chen_2023,Wang_2023}. These are generally characterized by slow dynamics and reduced transport due to the scarce connectivity between different many-body states~\cite{Fac4,Gromov_2020,Singh_2021,Richter_2022}. Physical manifestations of such models can be efficiently implemented in Rydberg quantum simulators, in which trapped atoms, excited to high-lying electronic states, feature strong state-dependent dipolar interactions~\cite{Saffmann2010,Adams_2019,Browaeys2020,Wu_2021}. Thanks to their versatility, these experimental platforms have led to several breakthroughs in the fields of quantum simulation and quantum computation~\cite{Jaksch2000,Bloch2012,Yang_Wang2016,Han_2016,Su2016,Petrosyan2017,Kaufman2021,Cohen2021,Pagano2022}. Concomitant to the strong electrostatic interactions are mechanical forces, that couple the internal atomic degrees of freedom to the external motional ones~\cite{Mehaignerie_2023,Qian_2023}. On the one hand, these forces can --- when uncontrolled --- be sources of undesired incoherent effects, such as dissipation and heating~\cite{Faoro2016,Robicheaux2021,Zhang2022}. On the other hand, coherent spin-phonon couplings allow to engineer long-range multi-body interactions~\cite{Gambetta2020}, to implement cooling protocols~\cite{Belyansky2019}, to explore polaron physics~\cite{Camargo_2018,Plodzien2018,Mazza2020,Magoni_phonon_dressing,Kosior_2023} and to realize artificial molecular systems~\cite{Gambetta2021,magoni_jt}. The impact of coherent lattice vibrations on the non-equilibrium dynamics of kinetically constrained quantum systems is currently unexplored. However, with the current advancements in the domain of Rydberg quantum simulation platforms, such studies will be soon within reach~\cite{Morgado_2021}.

In this work we explore the dynamics of elementary degrees of freedom (spin domains) in a chain of Rydberg atoms subject to the facilitation (anti-blockade) constraint. We show that interaction with the lattice vibrations manifests in an alteration of the power-law exponent characterizing the expansion of spin domains. At zero temperature the exponent does not depend on the spin-phonon coupling strength, provided that it is sufficiently weak, as scattering is off-resonant. However, when vibrational excitations are initially present, resonant scattering between phonons and spin domains leads to a quantitative change of the exponent. Coherent spin-phonon interactions thus may inhibit excitation transport, thereby providing a connection to disorder-free settings that display localization phenomena~\cite{Schiulaz_2015,Smith_2017}. Finally, we show that the spin domain expansion dynamics is sensitive to the phase of the vibrational states, which in translationally invariant systems may cause an asymmetric expansion~\cite{Valencia-Tortora_2023,Kitson_2023}.

\begin{figure}[ht]
    \centering
    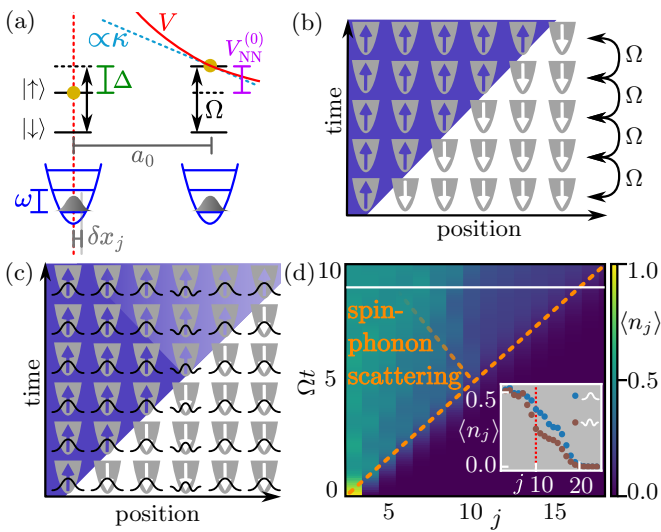
    \caption{\textbf{Spin-phonon scattering in a Rydberg chain.}
    (a) Atoms are treated as (fictitious) spins where the spin-down state is the ground state and the spin-up state is the Rydberg state. A laser with Rabi frequency $\Omega$ and detuning $\Delta$ excites the atoms to the Rydberg state under the facilitation condition $\Delta+V_\mathrm{NN}^{(0)}=0$, i.e., the detuning is chosen such that it cancels the nearest neighbor interaction $V_\mathrm{NN}^{(0)}$. Atoms are trapped in a state-independent harmonic potential with frequency $\omega$ and $\delta x_j$ is the deviation of the $j$-th atom from the center of the respective trap. Spin and motional degrees of freedom are coupled, with spin-phonon coupling constant $\kappa$ which is proportional to the gradient of the potential $V$ evaluated at the lattice spacing $a_0$.
    (b) Ballistic expansion of a spin domain (blue) in the absence of spin-phonon coupling: an excited Rydberg atom facilitates the excitation of the neighboring one at a rate proportional to $\Omega$. 
    (c) Spin-phonon scattering: the vibrational state of each atom is prepared in the ground state (Gaussian profile), except for one atom which is initialized in a higher lying Fock state. When the domain wall reaches this site, it scatters off the phonon excitation.
    (d) Numerical simulation of the spin-phonon scattering, where the atom at site $j=10$ is initialized in the Fock state $\ket{2}$ and all others in Fock state $\ket{0}$. Scattering (back-reflection) of the domain wall reduces the Rydberg density $\ev{n_j}$ beyond $j=10$ (orange dashed line). This is clearly seen in the inset which shows the Rydberg density $\ev{n_j}$ at time $\Omega t=9$ (white line in main plot) with the phonon excitation at site $j=10$ in Fock state $\ket{2}$ (brown points) and without it (blue points).
    }
    \label{fig:model}
\end{figure}

\textbf{Model ---}
We consider a one-dimensional chain of $N$ atoms, each one loaded in an optical tweezer trap and whose electronic structure is modeled as a two-level system (see Fig. \ref{fig:model}a). The state $\ket{\downarrow}$ denotes the ground state, while $\ket{\uparrow}$ represents the Rydberg (excited) state. The traps, which are separated by a nearest neighbour distance $a_0$, have a trap frequency $\omega$. The atoms are driven by a laser with Rabi frequency $\Omega$ and detuning $\Delta$ which couples the ground state to the Rydberg state. Two atoms in the Rydberg state, placed at sites $j$ and $k$, interact via a distance dependent potential of the form $V(\bm{r}_j,\bm{r}_k) = C_\gamma |\bm{r}_j - \bm{r}_k|^{-\gamma}$. Here $\gamma$ is the characteristic power law exponent (dipole-dipole interaction: $\gamma=3$, van der Waals interaction: $\gamma=6$).
The Hamiltonian of the system is then given by
\begin{equation*}
H = \sum_{j=1}^N \left( \Omega \sigma_j^x + \Delta n_j + \sum_{k < j} V(\bm{r}_j,\bm{r}_k) n_j n_k + \omega a_j^{\dagger} a_j  \right),
\label{eq:initial_Ham}
\end{equation*}
where $\sigma^x = \ket{\uparrow} \bra{\downarrow} + \ket{\downarrow} \bra{\uparrow}$ is the spin flip operator and $n = \ket{\uparrow} \bra{\uparrow}$ projects onto the Rydberg state. 
Writing the position fluctuations in the traps in terms of the bosonic operators as $\delta x_j = (a_j^{\dagger} + a_j)/\sqrt{2 m \omega}$ and neglecting the interactions beyond the nearest-neighbor ones, yields the simplified Hamiltonian \cite{Mazza2020,Magoni_phonon_dressing}
\begin{eqnarray}
    \label{eq:Ham_MPS}
    H &=& \sum_{j=1}^N \left\{ \Omega \sigma_j^x + \Delta n_j + \omega a_j^{\dagger} a_j \right. \\
    &&+ \left. \left[V_\mathrm{NN}^{(0)} - \kappa \left(a_j^{\dagger} + a_j - a_{j+1}^{\dagger} - a_{j+1} \right) \right] n_j n_{j+1}
     \right\}, \nonumber
\end{eqnarray}
where periodic boundary conditions are adopted. Here $V_\mathrm{NN}^{(0)}$ is the interaction between two excited nearest-neighboring atoms when they are located at the center of the respective traps and $\kappa=\gamma C_\gamma/(a_0^{\gamma+1} \sqrt{2m \omega})$
is the spin-phonon coupling constant, which is proportional to the gradient of the interaction potential evaluated at the lattice spacing $a_0$ (see Fig. \ref{fig:model}a). Such spin-phonon coupling accounts for the mechanical forces arising from the interaction between neighboring Rydberg excitations. These forces displace the atoms from the center of the respective traps only when they are in the Rydberg state, thereby coupling the internal (spin) degrees of freedom to the external (motional) ones.

\textbf{Facilitated dynamics ---}
We consider the situation in which the dynamics of the Rydberg chain is subject to the facilitation (anti-blockade) constraint \cite{Fac1,Amthor2010,Mattioli_2015,Su2017,Magoni_bloch_osc,Festa2022,Liu_2022}. This is obtained when the otherwise detuned laser is put on resonance by the single-atom energy shift induced by the Rydberg interaction, i.e., $\Delta + V_\mathrm{NN}^{(0)} = 0$. Under this condition, ground state atoms that are next to an already excited atom get resonantly coupled to the Rydberg state. The further assumption that both the next-nearest-neighbor interaction and the Rabi frequency are much smaller than the detuning, $V(\bm{r}_j^0,\bm{r}_{j+2}^0) \ll |\Delta|$ and $\Omega\ll|\Delta|$, leads to a constrained dynamics that conserves the number of domain walls delimiting domains of consecutive Rydberg excitations. This drastically reduces the connectivity between the many-body states, and the Hilbert space is decomposed into disconnected sectors, labeled by the number of domain walls~\cite{LocFac2}.

\begin{figure*}
    \centering
    \includegraphics[width=\linewidth]{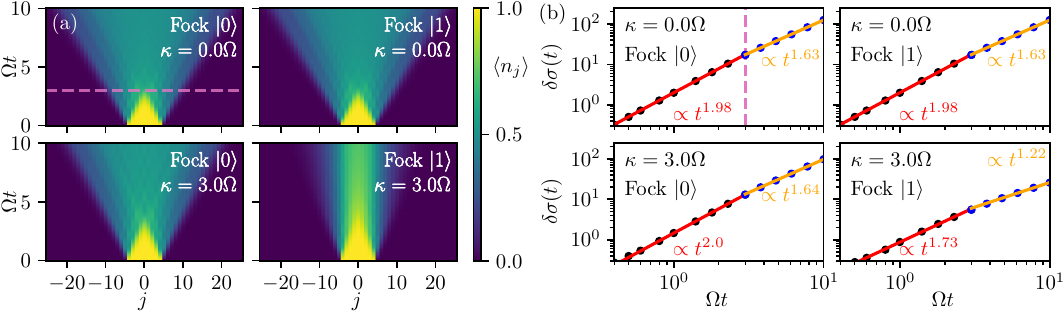}
    \caption{\textbf{Expansion of spin domain in the absence and presence of phonons.}
    (a) Rydberg density $\ev{n_j}$ for a spin domain initialized with $r_0=9$ Rydberg excitations and centered at $j=0$. In the absence of spin-phonon coupling ($\kappa=0$), the domain expands approximately ballistically, independently on whether the atoms are initially prepared in their vibrational ground state, $\ket{0}$ (left column), or first Fock state $\ket{1}$ (right column). When the spin-phonon coupling is switched on (we consider $\kappa=3.0\Omega$), ballistic expansion persists in the left column, while a drastic change is visible on the right. The reason is that when atoms are prepared in their vibrational ground state only virtual transitions to higher-lying phonon states take place, which merely renormalizes the ballistic propagation speed. In the presence of initial phonon excitations, however, coherent spin-phonon scattering takes place which alters the expansion dynamics dramatically.
    (b) The corresponding Rydberg density variance difference $\delta \sigma(t)$, Eq. (\ref{eq:increase_variance}). The power-law exponent changes at $\Omega t \approx 3$, where the initial domain wall has dissolved [see violet dashed line in the top left panel]. A clear change of exponent is observed in the bottom right panel ($\kappa = 3.0\Omega$, initial vibrational state of all atoms $\ket{1}$).
    Data is obtained via TEBD simulations of the dynamics under Hamiltonian (\ref{eq:Ham_MPS}) for $\omega=8\Omega,\Delta=500\Omega$. The maximum number of phonons per site is truncated to $7$.
    }
    \label{fig:Rydberg-facilitation}
\end{figure*}

Here, we focus on the single domain sector, i.e., the sector with two domain walls, and initially prepare a spin domain with $r_0$ consecutive Rydberg excitations. To evaluate its dynamics, we perform numerical simulations of Hamiltonian~\eqref{eq:Ham_MPS} utilizing the time-evolving block decimation algorithm (TEBD) \cite{TEBD,Orus2014,Gray2018,Paeckel2019,Cirac2021,Johansson2012}. In absence of spin-phonon coupling, i.e. $\kappa=0$, the two domain walls propagate freely along the lattice as free fermions~\cite{Ostmann_2019}. This results in a ballistic expansion of the spin domain whose size increases linearly in time, as sketched in Fig. \ref{fig:model}(b). To quantitatively assess this expansion, we evaluate the dynamics of the Rydberg density variance $\sigma$, defined as
\begin{equation}
\sigma(t) = \sum_{j=1}^N j^2 \frac{\ev{n_j(t)}}{\mathcal{N}(t)} - \left(\sum_{j=1}^N j \frac{\ev{n_j(t)}}{\mathcal{N}(t)} \right)^2, 
\label{eq:variance}
\end{equation}
where $\ev{\dots}$ denotes the quantum expectation value and $\mathcal{N}(t) = \sum_{k=1}^N \ev{n_k(t)}$ is the total Rydberg density.
The density variance~\eqref{eq:variance}, which can be measured experimentally~\cite{Ebadi2021}, quantifies the spreading dynamics of the spin domain.
It is connected to the mean square displacement used in Refs.~\cite{Steinigeweg_2017,Lezama_2022,Sierant_2023} and to the width of the density propagator studied in Ref.~\cite{Bera_2017} for a disordered fermionic model.
The density variance is expected to increase over time as
\begin{equation}
\delta \sigma(t) = \sigma(t) - \sigma(0) \sim t^\beta,
\label{eq:increase_variance}
\end{equation}
where $\sigma(0) = (r_0^2-1)/12$ is the density variance of the spin domain at $t=0$. When $\kappa = 0$, i.e. in the absence of spin-phonon coupling, we expect $\beta = 2$ (free fermions). This is indeed the case, as shown in the first two panels of Fig.~\ref{fig:Rydberg-facilitation}(a)-(b) where we plot the time evolution of the Rydberg density and 
$\delta \sigma(t)$, respectively. We note a crossover time that separates two regions characterized by two different exponents. The short-time behavior provides $\beta \approx 1.98$ reproducing our expectation, while for larger times the exponent decreases to $\beta \approx 1.63$. This behavior is a consequence of the conservation of the number of domain walls: throughout the facilitation dynamics, the domain walls cannot coalesce and therefore are subject to a hard-core repulsive potential \cite{Magoni_bloch_osc}. This translates into the interruption of the ballistic expansion when the two domain walls are about to collide, which happens at $\Omega t \approx 3$. Note that this effect is exclusively due to the finite size of the initial spin domain. 

In the presence of spin-phonon coupling, $\kappa \neq 0$, we find that the expansion of the spin domain strongly depends on the initial state of the phonons. In particular, when the atoms are initially prepared in their vibrational ground state $\otimes_{j=1}^N \ket{0}$, the initial ballistic expansion is maintained and the effect of the spin-phonon coupling is limited to a renormalization of the expansion velocity~\cite{Mazza2020,Magoni_phonon_dressing}. On the contrary, when the atoms are initially prepared in the first Fock state $\otimes_{j=1}^N \ket{1}$, the expansion of the domain changes dramatically and the exponent of the Rydberg density variance drops to $\beta \approx 1.73$ and $\beta \approx 1.22$ before and after the crossover time (see Fig.\ref{fig:Rydberg-facilitation}).

\textbf{Effective model for the spin-phonon scattering ---} The reason for the alteration of excitation transport in the presence of initial phonon excitations is resonant spin-phonon scattering, for which we will construct an effective model. Since the facilitation dynamics conserves the number of domain walls, the single spin domain can expand or shrink, but it is not allowed to split into two domains or disappear. The state of such spin domain can therefore be characterized with only two coordinates, namely its center of mass (CM) position and its relative coordinate (or the number of excitations it contains). The introduction of these two coordinates is particularly advantageous because it allows to reduce the complex many-body dynamics to a simpler two-body dynamics. As outlined in the Supplemental Material~\cite{supplement}, we formulate Hamiltonian~\eqref{eq:Ham_MPS} in terms of these two coordinates. By further decomposing the CM coordinate and the boson operator $a_j$ into Fourier modes respectively labeled by $q$ and $A_p$, and after applying various unitary transformations, one gets the Hamiltonian $H = \sum_{q=1}^N \ket q \bra q \otimes H_q$, with 
\begin{eqnarray}
\label{eq:H_q}
H_q &=& 2 J_q(\{N_p\}) \sum_{k=1}^{N-1} \cos\left(\frac{k \pi}{N}\right) \ket{k} \bra{k} + \omega \sum_p N_p \nonumber \\
&&+ \kappa \sum_{k,k^\prime,p} f_{k,k^\prime,p} \ket{k} \bra{k^\prime} \otimes \left(A_p + A_p^\dagger \right),
\end{eqnarray}
where $J_q(\{N_p\}) = 4 \Omega \cos\left[\frac{\pi}{N} \left(q + \sum_p p N_p \right) \right]$ and $N_p = A_p^\dagger A_p$. The first term provides a set of quasiparticle excitations labeled by their quasimomentum $k$, whose dispersion relation is connected to the expansion speed of the spin domain. Their interaction with the phonons is encoded in the third term, where $f_{k,k^\prime,p}$ are the spin-phonon coupling matrix elements which are derived in the Supplemental Material~\cite{supplement}. This spin-phonon coupling term is responsible for the change in the expansion of the spin domain shown in Fig.~\ref{fig:Rydberg-facilitation}. For $\kappa = 0$ the free dynamics of the quasiparticles results in the visible light cone emanating from the boundaries of the initial domain. In contrast, even for moderate values of $\kappa/\omega$, the presence of phonon excitations in the initial state drastically changes the dynamics of the domain. In order to analytically explore this regime, we note that the spin-phonon coupling term is the only one that does not conserve the total number of phonon excitations $N_\mathrm{phon} = \sum_p \ev{N_p}$. Therefore, when $|\kappa| \ll \omega$, the subspaces with different $N_\mathrm{phon}$ are only weakly coupled, making $N_\mathrm{phon}$ an approximately good quantum number. In this regime, we can derive an effective Hamiltonian that describes the facilitation dynamics in a given phonon subspace. This is formally accomplished by applying a Schrieffer-Wolff transformation~\cite{Schrieffer-Wolff} to Hamiltonian~\eqref{eq:H_q} so that we obtain an effective Hamiltonian, $H_\mathrm{eff}^{(q)}$, valid in each of the phonon subspaces, given by~\cite{supplement}
\begin{eqnarray}
H_\mathrm{eff}^{(q)} &=& 2 J_q(\{N_p\}) \sum_{k=1}^{N-1} \cos\left(\frac{k \pi}{N}\right) \ket{k} \bra{k} + \omega \sum_p N_p \nonumber \\
&&- \kappa^2 \sum_{k,k^\prime} F_{k,k^\prime}(\{A_p\}) \ket{k} \bra{k^\prime} + \mathcal{O}(\kappa^3).
\end{eqnarray}
This equation shows that the phonons mediate an effective interaction between quasiparticles, with matrix elements $F_{k,k^\prime}(\{A_p\})$. These contain terms like $A_m^\dagger A_n$, which have nonzero matrix elements only if $N_\mathrm{phon}>0$~\cite{supplement}. Therefore, when phonon excitations are already present in the initial state, these terms mediate additional interactions between quasiparticles that would not be present if the atoms were initialized in their vibrational ground state~\cite{Magoni_phonon_dressing}. This is consistent with the numerical results, shown in the bottom panels of Fig.~\ref{fig:Rydberg-facilitation}, that attribute the inhibition of the ballistic spin domain expansion to the presence of vibrational excitations in the initial state.  

\textbf{Phase sensitivity of spin-phonon scattering ---}
\begin{figure}
\centering
\includegraphics[width=\linewidth]{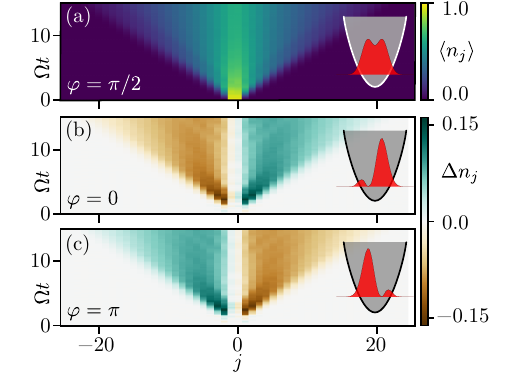}
\caption{\textbf{Phase sensitivity of spin-phonon scattering.}
(a) Expansion of the spin domain from the initial state with $r_0=2$ Rydberg excitations. All the atoms are initially prepared in the vibrational state $\ket{\varphi=\pi/2}$, giving rise to the position distribution in the trap shown in the right. The value of the Rydberg density $\ev{n_j}$ remains symmetric around the CM of the spin domain at all times. (b)-(c) Asymmetric domain expansion when the vibrational state of the atoms is initially prepared in $\ket{\varphi=0}$ and $\ket{\varphi=\pi}$ respectively. The asymmetry, which is quantified by $\Delta n_j = \ev{n_j}-\ev{n_{-j-1}}$, is caused by the asymmetry of the initial position distribution of the atoms (see inset). The simulations are carried out with $\omega=8\Omega,\Delta=200\Omega, \kappa=4\Omega$ and the maximum number of phonons per site is truncated to $3$.
}
\label{fig:phase-sensitivity}
\end{figure}
In the following we show that the coherence of spin-phonon scattering can be observed macroscopically. To this end we consider the situation in which the vibrational state of all the atoms is initially prepared in a coherent superposition of Fock states $\ket{0}$ and $\ket{1}$, as $\ket{\varphi} = 1/\sqrt{2} \left(\ket{0}+\mathrm{e}^{i\varphi}\ket{1}\right)$, which is specified by the phase $\varphi\in[0,2\pi)$.
Despite being initialized in a translationally invariant state, the spin domain expands generically in an asymmetric fashion around its initial position. This is seen in Fig.~\ref{fig:phase-sensitivity}. The asymmetry is controlled by the phase: for $\varphi = \pi/2$ the spin domain expands symmetrically, for $\varphi = 0$ ($\varphi = \pi$) the two domain walls propagate differently, with the right (left) front showing a larger Rydberg density. The emergence of this asymmetric expansion can be intuitively understood by the fact that for $\varphi \neq \pm \pi/2$, the probability distribution of the atomic position within each trap is not symmetric around its average value, as shown in the insets of Fig.~\ref{fig:phase-sensitivity}(b)-(c). This translates to the facilitation dynamics which fundamentally depends on the interatomic distance: the domain expands preferably into the direction in which the initial amplitude of the local vibrational wave packet is largest.

\textbf{Summary and outlook ---} 
We investigated the role of spin-phonon interaction on the non-equilibrium dynamics of Rydberg excitations in a chain of trapped atoms subject to the anti-blockade constraint. While aspects of our study are certainly idealized compared to the experimental state-of-the art, e.g. we assume state-independent trapping, we could identify coherent spin-phonon scattering as a mechanism that qualitatively alters the propagation of elementary excitations. In the future, it would be interesting to consider the impact of these processes in a many-body setting, by lifting the restriction to the single spin domain sector. Here one could ask whether the resulting complex spin-boson systems support the formation of localized many-body states in a disorder-free setting. Exploring this regime, which due to the huge Hilbert space size is challenging to treat on classical computers, could be an interesting use case for the next generation of quantum simulators.

\acknowledgments
\textit{Acknowledgments --- } We would like to thank T. Macrì for helpful discussions. We are grateful for financing from the Baden-Württemberg Stiftung through Project No. BWST\_ISF2019-23. We also acknowledge funding from the Deutsche Forschungsgemeinschaft (DFG, German Research Foundation) under Projects No. 428276754 and No. 435696605 as well as through the Research Unit FOR 5413/1, Grant No. 465199066 and the Research Unit FOR 5522/1, Grant No. 499180199. This project has also received funding from the European Union’s Horizon Europe research and innovation program under Grant Agreement No. 101046968 (BRISQ).
The authors acknowledge support by the state of Baden-Württemberg through bwHPC
and the German Research Foundation (DFG) through grant no INST 40/575-1 FUGG (JUSTUS 2 cluster). MM acknowledges support by the European Research Council (Grant No. 101041435).

\bibliography{bib}

\setcounter{equation}{0}
\setcounter{figure}{0}
\setcounter{table}{0}
\makeatletter
\renewcommand{\theequation}{S\arabic{equation}}
\renewcommand{\thefigure}{S\arabic{figure}}

\makeatletter
\renewcommand{\theequation}{S\arabic{equation}}
\renewcommand{\thefigure}{S\arabic{figure}}


\clearpage

\onecolumngrid
\setcounter{equation}{0}
\setcounter{figure}{0}
\setcounter{table}{0}
\setcounter{page}{1}

\begin{center}
{\Large SUPPLEMENTAL MATERIAL}
\end{center}
\begin{center}
\vspace{0.8cm}
{\Large Coherent spin-phonon scattering in facilitated Rydberg lattices}
\end{center}
\begin{center}
Matteo Magoni$^{1,2,3}$, Chris Nill$^{3,4}$, and Igor Lesanovsky$^{3,5}$
\end{center}
\begin{center}
$^1${\em Institute for Theoretical Physics, University of Innsbruck, Innsbruck 6020, Austria}\\
$^2${\em Institute for Quantum Optics and Quantum Information, Austrian Academy of Sciences, Innsbruck 6020, Austria}\\
$^3${\em Institut f\"ur Theoretische Physik, Universit\"at T\"ubingen, Auf der Morgenstelle 14, 72076 T\"ubingen, Germany}\\
$^4${\em{Institute for Applied Physics, University of Bonn, Wegelerstraße 8, 53115 Bonn, Germany}}\\
$^5${\em School of Physics and Astronomy and Centre for the Mathematics and Theoretical Physics of Quantum Non-Equilibrium Systems, The University of Nottingham, Nottingham, NG7 2RD, United Kingdom}
\end{center}

We derive in detail the results presented in the main text. The Supplemental Material is divided in two parts. In the first section, we derive Hamiltonian~(4) of the main text starting from the full many-body problem. In the second section, we explicitly apply the Schrieffer-Wolff transformation to obtain the effective Hamiltonian given by Eq.~(5) of the main text.

\section{Derivation of Hamiltonian~(4)}

In this section we show how to derive the Hamiltonian given by Eq.~(4) of the main text starting from the many-body Hamiltonian~(1) with the facilitation constraint $\Delta + V_\mathrm{NN}^{(0)} = 0$. The derivation goes through some intermediate steps. First, we reformulate the many-body problem in terms of three relevant coordinates where the state of the system can be written as Eq.~\eqref{sm:psi}. Second, we apply various unitary transformations to get Hamiltonian~\eqref{sm:full_ham}. Then, we make a change of basis to obtain Eq.~\eqref{sm:after_unitary} which coincides with Hamiltonian~(4) of the main text. Finally, we explicitly compute the spin-phonon coupling matrix elements in Eq.~\eqref{sm:spin_ph_matr_el}.

Consider a chain of atoms driven by a laser with Rabi frequency $\Omega$ and detuning $\Delta$. Neighboring Rydberg excited atoms interact with nearest-neighbor potential $V_\mathrm{NN}^{(0)}$. Under the facilitation constraint $\Delta + V_\mathrm{NN}^{(0)} = 0$, an atom in the ground state can be excited to the Rydberg state if it has only one neighboring atom that is already excited to the Rydberg state. This restrictive condition leads to a dynamics that conserves the total number of domain walls delimiting consecutive Rydberg excitations. As a consequence, the Hilbert space gets decomposed in many disconnected sectors, labeled by the number of domain walls. 

In our work we focus on the sector with a single spin domain, i.e., the one where two domain walls are present. Due to the conservation of the number of domain walls, the spin domain can only expand or shrink, but it cannot split or disappear. It is therefore possible to describe the state of the domain in terms of two coordinates, the position of its center of mass (CM) $c$ and its relative coordinate $r$, which is also equal to the number of Rydberg excitations. It is important to note that these two coordinates are not completely independent. Indeed, domains with an odd number of excitations have a CM located on a lattice site, while domains with an even number of excitations have a CM located on a midpoint between two lattice sites. These are the only two possibilities to have an admissible domain state. To explicitly distinguish between the two subsets of states, it is convenient to introduce a third coordinate $\xi = o,e$, which labels the parity of the number of excitations, i.e., $o$ for odd and $e$ for even. The state of the system can then be expressed as 
\begin{equation}
    \ket{\psi} = \ket c \otimes \ket r \otimes \ket{\xi},
\label{sm:psi}
\end{equation}
where $c = 1, 2 , \dots, N$, $r = 1, 2, \dots, \frac{N-1}{2}$ and $\xi = o,e$. Without loss of generality we assume that $N$ is an odd number. When $\xi = o$, for example, the spin domain has its CM on a lattice site and therefore the coordinate $c$ labels the integer CMs from $1$ to $N$, while $r$ labels the $\frac{N-1}{2}$ possible odd numbers between 1 and $N-1$. On the other hand, when $p=e$, the coordinate $c$ labels the $N$ half-integer CMs between $1/2$ and $N-1/2$, while $r$ labels the $\frac{N-1}{2}$ possible even numbers between 1 and $N-1$. For instance, according to this notation and denoting ground state atoms with $\ket{\downarrow}$ and Rydberg atoms with $\ket{\uparrow}$, $\ket{\uparrow \uparrow \uparrow \downarrow \downarrow \dots} = \ket{2,2,o}$ and $\ket{\downarrow \uparrow \uparrow \downarrow \downarrow \dots} = \ket{3,1,e}$ are two admissible domain states. 

Given this representation and assuming periodic boundary conditions, under the facilitation constraint, a state $\ket{c, r, o}$ is resonant with only four other states, provided that $1 < r < N-1$ (when $r=1$ the spin domain can only increase, when $r=N-1$ it can only decrease). These are: $\ket{c,r,e}$ (the spin to the left of the leftmost excitation flips up), $\ket{c+1, r,e}$ (the spin to the right of the rightmost excitation flips up), $\ket{c+1,r-1,e}$ (the leftmost excitation flips down), $\ket {c,r-1,e}$ (the rightmost excitation flips down). On the other hand, a state $\ket{c, r, e}$ is resonantly coupled with $\ket{c-1,r+1,o}$, $\ket{c,r+1,o}$, $\ket{c,r,o}$, $\ket{c-1,r,o}$. 

We further introduce the operators $a_j$ and $a_j^\dagger$ as bosonic annihilation and creation operators, defined with respect to the displacement of the position $\bm{r}_j$ of the $j$-th atom, from the center of the respective trap $\bm{r}_j^0$. The traps are separated by a nearest neighbor distance $a_0$. Assuming that $|\delta \bm{r}_j| \ll a_0$, the potential around the equilibrium positions can be approximated to leading order as $V(\bm{r}_j,\bm{r}_k) \simeq V(\bm{r}_j^0, \bm{r}_k^0) + \nabla  V(\bm{r}_j,\bm{r}_k) | _{(\bm{r}_j^0,\bm{r}_k^0)} \cdot (\delta \bm{r}_j, \delta \bm{r}_k)$, where the expansion needs to be done only along the $x$-direction, i.e., parallel to the chain.
Writing the atom fluctuations in terms of the bosonic operators as $\delta x_j = (a_j^{\dagger} + a_j)/\sqrt{2 m \omega}$, neglecting the interactions beyond the nearest-neighbor ones and using the representation in terms of the CM and relative coordinates, we can write the Hamiltonian of a single domain of consecutive Rydberg excitations as ($\hbar = 1$)
\begin{eqnarray}
H  = & 2 \Omega &\sum_{c=1}^N \sum_{r=2}^\frac{N-1}{2} \left[\left(\ket{c}+ \ket{c+1}\right) \bra c  \otimes \left(\ket{r} + \ket{r-1} \right) \bra{r} \otimes \ket{e}\bra{o} + \mathrm{h.c.} \right] \nonumber \\
& -& \kappa \sum_{c=1}^N \sum_{r=2}^\frac{N-1}{2} \ket c \bra c \otimes \ket r \bra r \otimes \ket o \bra o \otimes \left(a_{c+r-1} - a_{c-r+1} + \mathrm{h.c.} \right) \nonumber \\
& -& \kappa \sum_{c=1}^N \sum_{r=1}^\frac{N-1}{2} \ket c \bra c \otimes \ket r \bra r \otimes \ket e \bra e \otimes \left(a_{c+r-1} - a_{c-r} + \mathrm{h.c.} \right) \nonumber \\
& +& \omega \sum_{j=1}^N a_j^{\dagger} a_j.
\label{sm:hamilt}
\end{eqnarray} 
The first term of Eq.~\eqref{sm:hamilt} is the kinetic energy of the spin domain and couples states in the $o/e$ sector to states in the $e/o$ sector. The second and third terms contain the coupling between the degrees of freedom of the spin domain and the bosons in the odd and even sector respectively. Note that the subscripts of the boson operators refer to the physical lattice and are written in terms of $c$ and $r$ differently in the two cases. The coupling constant is
\begin{equation}
 \kappa = \frac{\gamma\,C_\gamma}{a_0^{\gamma+1} \sqrt{2 m \omega}} =\frac{\gamma}{\sqrt{2}}\frac{x_0}{a_0}V_\mathrm{NN}^{(0)}
\end{equation}
and depends on microscopic quantities, like the gradient of the interaction potential evaluated at the interatomic distance (which for the power-law potential considered here can be expressed in terms of the nearest-neighbor interaction $V_\mathrm{NN}$) and the harmonic oscillator length $x_0 = \sqrt{1/(m \omega)}$.

We proceed by introducing the Fourier modes of the boson operators through $a_j = \frac{1}{\sqrt{N}} \sum_p e^{i(2\pi/N) j p} A_p$, with $p=1, 2 \dots, N$. In terms of the operators $A_p$, the Hamiltonian reads
\begin{eqnarray}
H  = & 2 \Omega &\sum_{c=1}^N \sum_{r=1}^\frac{N-1}{2} \left[\left(\ket{c}+ \ket{c+1}\right) \bra c  \otimes \left(\ket{r} + \ket{r-1} \right) \bra{r} \otimes \ket{e}\bra{o} + \mathrm{h.c.} \right] \nonumber \\
& -& \frac{\kappa}{\sqrt{N}} \sum_{c=1}^N \sum_{r=1}^\frac{N-1}{2} \ket c \bra c \otimes \ket r \bra r \otimes \left \{ \ket o \bra o \otimes \left[2i \sum_p e^{i(2\pi/N)cp} \sin{\left[\frac{2\pi}{N}(r-1)p \right]} A_p + \mathrm{h.c.} \right] \right. \nonumber \\
&+& \left. \ket e \bra e \otimes \left[2i \sum_p e^{i (2\pi/N) \left(c-\frac{1}{2} \right)p} \sin{\left[\frac{2\pi}{N}\left(r-\frac{1}{2} \right)p \right]} A_p + \mathrm{h.c.} \right] \right \} + \omega \sum_{j=1}^N N_p,
\label{sm:ham_after_fourier}
\end{eqnarray}
where $N_p = A_p^{\dagger} A_p$. The CM degree of freedom can be now decoupled from the boson modes by applying the so-called Lee-Low-Pines (LLP) transformation~\cite{LLP_transf}
\begin{equation*}
U_\mathrm{LLP} = \left\{\exp\left[-i \sum_c c \ket{c} \bra{c} \otimes \sum_p \frac{2\pi}{N} p N_p\right] \otimes \ket o \bra o + \exp\left[-i \sum_c \left(c-\frac{1}{2} \right) \ket{c} \bra{c} \otimes \sum_p \frac{2\pi}{N} p N_p\right] \otimes \ket e \bra e \right\} \exp \left( - i \frac{\pi}{2} \sum_p N_p \right)
\end{equation*}
to Hamiltonian~\eqref{sm:ham_after_fourier}. By further introducing the Fourier decomposition of the CM coordinate, $c = \frac{1}{\sqrt{N}} \sum_q e^{i(2\pi/N)cq} \ket q$, where $q = 1,2, \dots, N$, the Hamiltonian is finally reduced to a block-diagonal form as $U_\mathrm{LLP}^\dagger H U_\mathrm{LLP} = \sum_q \ket q \bra q \otimes \widetilde{H}_q$. Hence, after the LLP and the Fourier transforms, the mode $q$ of the CM becomes a good quantum number, and the Hamiltonian $\widetilde{H}_q$ governing the evolution within a given $q$ sector is given by
\begin{eqnarray*}
\widetilde{H}_q &= & 4 \Omega  \cos\left[\frac{\pi}{N} \left(q + \sum_p p N_p \right) \right] \left\{e^{i \pi q/N} \sum_{r=1}^\frac{N-1}{2} \left(\ket{r} + \ket{r-1} \right) \bra r \otimes \ket e \bra o + \mathrm{h.c.} \right\} + \omega \sum_p N_p \\
&&- \frac{2 \kappa}{\sqrt{N}} \sum_{r=1}^\frac{N-1}{2} \ket{r} \bra{r} \otimes \left\{ \ket o \bra o \otimes \sum_p \sin\left[\frac{2\pi}{N}(r-1) p \right] \left(A_p + A_p^{\dagger}\right) + \ket e \bra e \otimes \sum_p \sin\left[\frac{2\pi}{N} \left(r-\frac{1}{2} \right) p \right] \left(A_p + A_p^{\dagger}\right) \right\}. \nonumber
\end{eqnarray*}
By further applying the unitary
\begin{equation*}
    U_q = e^{-i \pi q/N} \ket{o} \bra{o} + \ket{e} \bra{e}
\end{equation*}
and making rid of the parity coordinate $\xi=e,o$ through the mapping
\begin{eqnarray*}
    &&\ket{r,e} \longleftrightarrow \ket{2r} \nonumber \\
    &&\ket{r,o} \longleftrightarrow \ket{2r-1},
\end{eqnarray*}
one obtains the Hamiltonian $H_q^\prime = U_q^\dagger \widetilde{H}_q U_q$ given by
\begin{eqnarray}
H_q^\prime &= & J_q(\{N_p\}) \sum_{r^\prime=1}^{N-2} \left( \ket{r^\prime+1} \bra {r^\prime} + \mathrm{h.c.} \right) + \omega \sum_p N_p \nonumber \\
&&- \frac{2 \kappa}{\sqrt{N}} \sum_{r^\prime=1}^{N-1} \ket{r^\prime} \bra{r^\prime} \sum_p \sin\left[\frac{\pi}{N}\left(r^\prime-1\right) p \right] \left(A_p + A_p^{\dagger}\right),
\label{sm:full_ham}
\end{eqnarray}
where $J_q(\{N_p\}) = 4 \Omega \cos\left[\frac{\pi}{N} \left(q + \sum_p p N_p \right) \right]$ and $r^\prime = 2r$. This Hamiltonian is composed of three terms. The first is a hopping term in the relative coordinate which represents the shrinking and expansion of the spin domain under the facilitation constraint, with hopping amplitude $J_q(\{N_p\})$ that depends on the phonon (momentum) occupation number. The second term is the kinetic energy of the phonons. The third is a coupling term between the dynamics of the relative coordinate and the phonon dynamics.

This Hamiltonian has the same form of the effective Hamiltonian derived in Ref.~\cite{Magoni_phonon_dressing}, with the only difference that now the first Brillouin zone of the CM modes is halved. The present derivation is more accurate because the domain of Hamiltonian~\eqref{sm:hamilt} is the Hilbert space that contains states of the form~\eqref{sm:psi}, which are the only spin domains that are allowed to exist. In Ref.~\cite{Magoni_phonon_dressing} the Hamiltonian is instead defined on a larger Hilbert space that also contains states that are not admissible, like the ones that have the CM on a lattice site and even number of excitations. However, these ``illegal'' states are only coupled to other illegal states, while admissible states are only coupled to other admissible states. This leads to the presence of two disconnected sectors that are characterized by the same dynamics and therefore possess the same spectrum. For this reason in Ref.~\cite{Magoni_phonon_dressing} the first Brillouin zone of the CM modes is twice than the one in this paper. However, the dynamics from an admissible initial state is identical in both treatments. 

To get further insights, we rewrite Hamiltonian~\eqref{sm:full_ham} in the basis that simultaneously diagonalizes its first two terms. This is accomplished by applying the unitary 
\begin{equation}
U = \sqrt{\frac{2}{N}} \sum_{k,j = 1}^{N-1} \sin\left(\frac{\pi}{N} k j \right) \ket{j} \bra{k} \otimes \mathds{1}_\mathrm{bos},
\label{s_eq:unitary}
\end{equation}
where $\mathds{1}_\mathrm{bos}$ stands for the identity operator on the bosonic Fock space. We therefore get the Hamiltonian $H_q = U^\dagger H_q^\prime U$ given by
\begin{eqnarray}
H_q &=& 2 J_q(\{N_p\}) \sum_{k=1}^{N-1} \cos\left(\frac{k \pi}{N}\right) \ket{k} \bra{k} + \omega \sum_p N_p \nonumber \\  
&&- U^\dagger \left\{\frac{2 \kappa}{\sqrt{N}} \sum_{r=1}^{N-1} \ket{r} \bra{r} \sum_p \sin\left[\frac{\pi}{N}\left(r-1\right) p \right] \left(A_p + A_p^{\dagger}\right) \right\} U,
\label{sm:after_unitary}
\end{eqnarray}
which coincides with Hamiltonian~(4) in the main text, except for the last term that corresponds to the spin-phonon scattering and where the application of the unitary transformation needs to be explicitly evaluated. In order to evaluate it, using Eq.~\eqref{s_eq:unitary} and keeping the constant factor $-4 \kappa N^{-3/2}$ out, we can rewrite the second line of Eq.~\eqref{sm:after_unitary} as
\begin{eqnarray*}
\sum_{k,j,k^\prime,j^\prime,r,p} \sin\left(\frac{\pi}{N} k j \right) \ket{k} \bra{j} \sin\left[\frac{\pi}{N}\left(r-1\right) p \right] \ket{r} \bra{r} \sin\left(\frac{\pi}{N} k^\prime j^\prime \right) \ket{j^\prime} \bra{k^\prime} \otimes \left(A_p + A_p^{\dagger}\right) \nonumber \\ 
= \sum_{k,k^\prime,r,p} \sin\left(\frac{\pi}{N} k r \right)  \sin\left[\frac{\pi}{N} p (r-1) \right] \sin\left(\frac{\pi}{N} k^\prime r \right) \ket{k} \bra{k^\prime} \otimes \left(A_p + A_p^\dagger \right),
\end{eqnarray*}
where the sums over $k,j,k^\prime,j^\prime,r$ run from 1 to $N-1$, the sum over $p$ runs from 1 to $N$ and we make use of the orthonormality of the basis states. We therefore need to compute the coefficients
\begin{eqnarray}
\label{s_eq:coefficients}
\frac{f_{k,k^\prime,p}}{-4 N^{-3/2}} &=& \sum_{r=1}^{N-1} \sin\left(\frac{\pi}{N} k r \right)  \sin\left[\frac{\pi}{N} p (r-1) \right] \sin\left(\frac{\pi}{N} k^\prime r \right) \\ \nonumber
&=& \cos\left(\frac{\pi}{N}p \right) \sum_{r=1}^{N-1} \sin\left(\frac{\pi}{N} k r \right)  \sin\left(\frac{\pi}{N} p r \right) \sin\left(\frac{\pi}{N} k^\prime r \right) - \sin\left(\frac{\pi}{N} p \right) \sum_{r=1}^{N-1} \sin\left(\frac{\pi}{N} k r \right)  \cos\left(\frac{\pi}{N} p r \right) \sin\left(\frac{\pi}{N} k^\prime r \right).
\end{eqnarray}
The two terms can be evaluated separately using the Werner's formulae, which decompose products of trigonometric functions into sums of trigonometric functions. For example, the first sum in Eq.~\eqref{s_eq:coefficients} can be rewritten as
\begin{eqnarray}
&&\sum_{r=1}^{N-1} \sin\left(\frac{\pi}{N} k r \right)  \sin\left(\frac{\pi}{N} p r \right) \sin\left(\frac{\pi}{N} k^\prime r \right) \nonumber \\
&&= \frac{1}{4} \sum_{r=1}^{N-1} \left\{\sin \left[\frac{\pi}{N}(k-k^\prime+p) r \right] - \sin \left[\frac{\pi}{N}(k-k^\prime-p) r \right] - \sin \left[\frac{\pi}{N}(k+k^\prime+p) r \right] + \sin \left[\frac{\pi}{N}(k+k^\prime-p) r \right] \right\} \nonumber \\
&&= \begin{cases}
- \frac{\sin\left(\frac{\pi}{N}p\right) \sin\left(\frac{\pi}{N}k\right) \sin\left(\frac{\pi}{N}k^\prime \right)}{\cos^2 \left(\frac{\pi}{N}p\right) + \cos^2 \left(\frac{\pi}{N}k\right) + \cos^2 \left(\frac{\pi}{N}k^\prime\right) - 2 \cos \left(\frac{\pi}{N}p\right) \cos \left(\frac{\pi}{N}k\right) \cos \left(\frac{\pi}{N}k^\prime \right) - 1} \qquad &\text{for odd} \; k-k^\prime-p \\
0 \qquad &\text{for even} \; k-k^\prime-p
\end{cases},
\label{s_eq:first_term}
\end{eqnarray}
where from the second to the third line we make use of the sum
\begin{equation*}
\sum_{r=1}^{N-1} \sin\left(\frac{\pi}{N} \alpha r \right) = 
\begin{cases} \frac{\sin \left(\frac{\pi}{N} \alpha \right)}{1-\cos \left(\frac{\pi}{N} \alpha \right)} \qquad &\text{for odd} \; \alpha \\
0 \qquad &\text{for even} \; \alpha 
\end{cases}.
\end{equation*}
Analogously, the second sum in Eq.~\eqref{s_eq:coefficients} can be evaluated as
\begin{eqnarray}
&&\sum_{r=1}^{N-1} \sin\left(\frac{\pi}{N} k r \right)  \cos\left(\frac{\pi}{N} p r \right) \sin\left(\frac{\pi}{N} k^\prime r \right) \nonumber \\
&&= \frac{1}{4} \sum_{r=1}^{N-1} \left\{\cos \left[\frac{\pi}{N}(k-k^\prime+p) r \right] + \cos \left[\frac{\pi}{N}(k-k^\prime-p) r \right] - \cos \left[\frac{\pi}{N}(k+k^\prime+p) r \right] - \cos \left[\frac{\pi}{N}(k+k^\prime-p) r \right] \right\} \nonumber \\
&&= -\frac{N}{4} \left[\left(\delta_{k-k^\prime,-p} + \delta_{k-k^\prime,p} \right) - \left(\delta_{k+k^\prime,2N-p} + \delta_{k+k^\prime,p} \right) \right],
\label{s_eq:second_term}
\end{eqnarray}
where we make use of the formula
\begin{equation*}
\sum_{r=1}^{N-1} \cos\left(\frac{\pi}{N} \alpha r \right) 
= \begin{cases}
0 \qquad &\text{for odd} \; \alpha \\
-1 \qquad &\text{for even} \; \alpha \neq 2mN \\
N-1 \qquad &\text{for} \; \alpha = 2mN 
\end{cases},
\end{equation*}
with $m \in \mathds{Z}$.

Inserting the two terms~\eqref{s_eq:first_term} and~\eqref{s_eq:second_term} back into Eq.~\eqref{s_eq:coefficients} yields the spin-phonon coupling matrix elements
\begin{eqnarray}
\label{sm:spin_ph_matr_el}
f_{k,k^\prime,p} = 4 N^{-3/2} & \left \{\cos\left(\frac{\pi}{N} p \right) \left[\frac{\sin\left(\frac{\pi}{N}p\right) \sin\left(\frac{\pi}{N}k\right) \sin\left(\frac{\pi}{N}k^\prime \right)}{\cos^2 \left(\frac{\pi}{N}p\right) + \cos^2 \left(\frac{\pi}{N}k\right) + \cos^2 \left(\frac{\pi}{N}k^\prime\right) - 2 \cos \left(\frac{\pi}{N}p\right) \cos \left(\frac{\pi}{N}k\right) \cos \left(\frac{\pi}{N}k^\prime \right) - 1} \right] \delta_{k-k^\prime-p, \mathrm{odd}} \right. \nonumber \\ 
& \left. + \frac{N}{4} \sin\left(\frac{\pi}{N} p \right) \left[\left(\delta_{k-k^\prime,-p} + \delta_{k-k^\prime,p} \right) - \left(\delta_{k+k^\prime,2N-p} + \delta_{k+k^\prime,p} \right) \right] \right\}
\end{eqnarray}
present in Hamiltonian~(4) of the main text. These matrix elements are composed of two terms. The second term is connected to the momentum conservation in the scattering process between the quasiparticles $\ket{k}$ and the phonons with momentum $p$. The first term is nonzero whenever the quantity $k-k^\prime-p$ is an odd number. The elaborate structure of the matrix elements $f_{k,k^\prime,p}$ originates from the complex scattering mechanism of a composite object, i.e., the domain of Rydberg excitations, with the phonons.

\section{Derivation of the effective Hamiltonian~(5)} 

In this section we derive the effective Hamiltonian given by Eq.~(5) in the main text, which is valid for $|\kappa| \ll \omega$. The idea is to start from Hamiltonian~(4) of the main text and perform a Schrieffer-Wolff transformation to trace out the phonon degrees of freedom. The standard procedure is to split the Hamiltonian $H = H_0 + V$ into a diagonal part $H_0$ and a small off-diagonal perturbation $V$. The Schrieffer-Wolff transformation is performed by applying a unitary transformation with generator $S$ to the Hamiltonian as
\begin{equation}
\label{sm:SW_transform}
H^\prime = e^S H e^{-S} = H + [S,H] + \frac{1}{2}[S,[S,H]] + \dots,
\end{equation}  
where we make use of the Baker-Campbell-Haussdorf formula. Since $V$ is small, the generator $S$ of the transformation is also small. In terms of $H_0$ and $V$, Eq.~\eqref{sm:SW_transform} can be rewritten as
\begin{equation*}
H^\prime = H_0 + V + [S,H_0] + [S,V] + \frac{1}{2}[S,[S,H_0]] + \frac{1}{2}[S,[S,V]] + \dots
\end{equation*}
which can be made diagonal to first order in $V$ by choosing the generator $S$ such that
\begin{equation}
\label{sm:SW_condition}
V + [S,H_0] = 0.
\end{equation}
In this way, the transformed Hamiltonian reads
\begin{equation*}
H^\prime = H_0 + \frac{1}{2}[S,V] + \mathcal{O}(V^3), 
\end{equation*}
which can finally be projected to the subspace of interest to derive an effective Hamiltonian for that subspace. 

In our case we identify (see Eq.~(4) of the main text)
\begin{equation*}
H_0 =  2 J_q(\{N_p\}) \sum_{k=1}^{N-1} \cos\left(\frac{k \pi}{N}\right) \ket{k} \bra{k} + \omega \sum_p N_p,
\end{equation*}
which is diagonal in the basis $\{\ket{k}\}_{k=1,\dots,N-1} \otimes \ket{N_1, \dots, N_N}$ with eigenvalue $E_{k,\{N_p\}} = 2 J_q(\{N_p\}) \cos\left(\frac{k \pi}{N}\right) + \omega \sum_p N_p$, and
\begin{equation*}
V = \kappa \sum_{k,k^\prime,p} f_{k,k^\prime,p} \ket{k} \bra{k^\prime} \otimes \left(A_p + A_p^\dagger \right),
\end{equation*}
which couples states whose total number of phonons differs by 1. We find that the generator $S$ that satisfies Eq.~\eqref{sm:SW_condition} is
\begin{equation*}
S = - \kappa \sum_{k,k^\prime,p} f_{k,k^\prime,p} \ket{k} \bra{k^\prime} \otimes \left(\frac{1}{d_1^{k,k^\prime,p}} A_p^\dagger + \frac{1}{d_2^{k,k^\prime,p}} A_p \right),
\end{equation*}
where the two denominators are energy differences: $d_1^{k,k^\prime,p} = E_{k^\prime,\{N_p\}} - E_{k,\{N_p+1\}}$ and $d_2^{k,k^\prime,p} = E_{k^\prime,\{N_p+1\}} - E_{k,\{N_p\}}$. According to this notation, $E_{k,\{N_p+1\}}$ is the energy of the state that has one phonon with momentum $p$ more than the state with energy $E_{k,\{N_p\}}$. Computing the commutator $\frac{1}{2}[S,V]$ provides
\begin{eqnarray*}
\frac{1}{2}[S,V] &= \frac{1}{2} \left(SV - VS \right) \nonumber \\
&= - \frac{\kappa^2}{2} \sum_{k,k^\prime,p,k^{\prime \prime}, k^{\prime \prime \prime}, p^\prime} & \left\{f_{k,k^\prime,p} f_{k^{\prime \prime}, k^{\prime \prime \prime}, p^\prime} \ket{k} \braket{k^{\prime}|k^{\prime \prime}} \bra{k^{\prime \prime \prime}} \left[\frac{1}{d_1^{k,k^\prime,p}} A_p^\dagger \left(A_{p^\prime} + A_{p^\prime}^\dagger \right) + \frac{1}{d_2^{k,k^\prime,p}} A_p \left(A_{p^\prime} + A_{p^\prime}^\dagger \right) \right] \right. \nonumber \\
&&- \left. f_{k^{\prime \prime}, k^{\prime \prime \prime}, p^\prime} f_{k,k^\prime,p} \ket{k^{\prime \prime}} \braket{k^{\prime \prime \prime}|k} \bra{k^{\prime}} \left[\frac{1}{d_1^{k,k^\prime,p}} \left(A_{p^\prime} + A_{p^\prime}^\dagger \right) A_p^\dagger + \frac{1}{d_2^{k,k^\prime,p}} \left(A_{p^\prime} + A_{p^\prime}^\dagger \right)  A_p \right] \right\} \nonumber \\
&= - \frac{\kappa^2}{2} \sum_{k,k^\prime,p, k^{\prime \prime \prime}, p^\prime} & \left\{f_{k,k^\prime,p} f_{k^{\prime}, k^{\prime \prime \prime}, p^\prime} \ket{k} \bra{k^{\prime \prime \prime}} \left[\frac{1}{d_1^{k,k^\prime,p}} A_p^\dagger A_{p^\prime} + \frac{1}{d_2^{k,k^\prime,p}} A_p A_{p^\prime}^\dagger \right] \right. \nonumber \\
&&- \left. f_{k^{\prime \prime}, k, p} f_{k,k^\prime,p^\prime} \ket{k^{\prime \prime}} \bra{k^{\prime}} \left[\frac{1}{d_1^{k,k^\prime,p}} A_{p^\prime} A_p^\dagger + \frac{1}{d_2^{k,k^\prime,p}} A_{p^\prime}^\dagger A_p \right] \right\} \nonumber \\
&= - \frac{\kappa^2}{2} \sum_{k,k^\prime,p, k^{\prime \prime \prime}, p^\prime} & f_{k,k^\prime,p} f_{k^{\prime}, k^{\prime \prime \prime}, p^\prime} \ket{k} \bra{k^{\prime \prime \prime}} \left[\left(\frac{1}{d_1^{k,k^\prime,p}} - \frac{1}{d_1^{k^\prime,k^{\prime \prime \prime},p}} \right) A_p^\dagger A_{p^\prime} + \left(\frac{1}{d_2^{k,k^\prime,p}} - \frac{1}{d_2^{k^\prime,k^{\prime \prime \prime},p}} \right) A_{p^\prime}^\dagger A_p \right. \nonumber \\
&&+ \left. \left(\frac{1}{d_2^{k,k^\prime,p}} - \frac{1}{d_1^{k^\prime,k^{\prime \prime \prime},p}} \right) \delta_{p,p^\prime} \right],
\end{eqnarray*}
where from the second to the third line we have neglected all the terms that do not conserve the total number of phonons. From the third to the fourth line we have renamed the indexes as $k^{\prime \prime} \rightarrow k$, $k \rightarrow k^{\prime}$ and $k^{\prime} \rightarrow k^{\prime \prime \prime}$ in the second part of the sum. Defining 
\begin{equation}
F_{k, k^{\prime \prime \prime}}(\{A_p\}) = \frac{1}{2} \sum_{k^\prime,p, p^\prime} f_{k,k^\prime,p} f_{k^{\prime}, k^{\prime \prime \prime}, p^\prime} \left[\left(\frac{1}{d_1^{k,k^\prime,p}} - \frac{1}{d_1^{k^\prime,k^{\prime \prime \prime},p}} \right) A_p^\dagger A_{p^\prime} + \left(\frac{1}{d_2^{k,k^\prime,p}} - \frac{1}{d_2^{k^\prime,k^{\prime \prime \prime},p}} \right) A_{p^\prime}^\dagger A_p + \left(\frac{1}{d_2^{k,k^\prime,p}} - \frac{1}{d_1^{k^\prime,k^{\prime \prime \prime},p}} \right) \delta_{p,p^\prime} \right],
\label{sm:F_k_k_third}
\end{equation}
one gets the effective interaction term between the quasiparticles that appears in Eq.~(5) of the main text. Note that Eq.~\eqref{sm:F_k_k_third} contains terms like $A_p^\dagger A_{p^\prime}$, which have nonzero matrix elements only if the Hamiltonian is projected onto a phonon subspace that contains phonon excitations. We attribute to these terms the qualitative change in the expansion of the spin domain when vibrational excitations are present in the initial state.




\end{document}